\documentclass[10pt,english]{IEEEtran}
\usepackage[T1]{fontenc}
\usepackage{geometry}
\usepackage{amsmath}
\usepackage{xpatch}
\usepackage{amssymb}
\usepackage{esint}
\usepackage{mathtools}
\usepackage{amsthm}
\usepackage{nicefrac}
\usepackage{gensymb}
\usepackage{bigints}
\usepackage{mathtools}
\usepackage{blkarray, bigstrut}
\usepackage{physics}
\usepackage{calligra}
\usepackage{graphicx}
\usepackage{epstopdf}
\usepackage{dsfont}
\usepackage{array,ragged2e}
\usepackage{enumitem}
\usepackage{etoolbox}
\usepackage{babel}
\usepackage[nopar]{lipsum}
\usepackage{psfrag}
\usepackage[ruled,norelsize] {algorithm2e}
\usepackage{algorithm2e}
\usepackage{balance}
\usepackage{float}
\usepackage{verbatim}
\usepackage[caption = false]{subfig}
\SetKw{KwBy}{by}
\usepackage{algpseudocode}
\usepackage{makecell}
\usepackage{to-be-determined} 

\usepackage[compact]{titlesec}

\titlespacing{\section}{0.5pt}{*0.5}{*0.5}
\titlespacing{\subsection}{0.5pt}{*0.5}{*0.5}
\titlespacing{\subsubsection}{0.5pt}{*0.5}{*0.5}

\makeatletter
\newcommand{\removelatexerror}{\let\@latex@error\@gobble}
\makeatother

\newtheoremstyle{plain}
  {\topsep}   
  {\topsep}   
  {\itshape}  
  {0pt}       
  {\bfseries} 
  {.}         
  {5pt plus 1pt minus 1pt} 
  {\thmname{#1}\thmnumber{ #2} \textnormal{(\thmnote{#3})}} 

\xpatchcmd{\proof}{\hskip\labelsep}{\hskip5\labelsep}{}{}  
\makeatletter
\xpatchcmd{\proof}{\@addpunct{.}}{\@addpunct{:}}{}{}
\makeatother

\renewcommand\[{\begin{equation}}
\renewcommand\]{\end{equation}} 
\usepackage{listings}
\usepackage{fancyvrb}
\usepackage{framed}

\usepackage{courier}
\definecolor{dkgreen}{rgb}{0,0.3,0}
\definecolor{gray}{rgb}{0.5,0.5,0.5}

\begin{document}

\title{Towards Bridging the FL Performance-Explainability Trade-Off: A Trustworthy 6G RAN Slicing Use-Case}
\author{Swastika Roy,~\IEEEmembership{Student Member,~IEEE}, Hatim Chergui,~\IEEEmembership{Senior Member,~IEEE}\\and Christos Verikoukis,~\IEEEmembership{Senior Member,~IEEE}\\
\IEEEcompsocitemizethanks{\IEEEcompsocthanksitem S. Roy is with Iquadrat Informatica and Universitat Politècnica de Catalunya (UPC), Barcelona, Spain, H. Chergui is with i2CAT Foundation, Barcelona, Spain and C. Verikoukis is with University of Patras and ISI/ATHENA, Greece. [e-mail: s.roy@iquadrat.com, chergui@ieee.org, cveri@isi.gr]}}

\maketitle
\thispagestyle{empty}

\begin{abstract}
In the context of sixth-generation (6G) networks, where diverse network slices coexist, the adoption of AI-driven zero-touch management and orchestration (MANO) becomes crucial. However, ensuring the trustworthiness of AI black-boxes in real deployments is challenging. Explainable AI (XAI) tools can play a vital role in establishing transparency among the stakeholders in the slicing ecosystem. But there is a trade-off between AI performance and explainability, posing a dilemma for trustworthy 6G network slicing because the stakeholders require both highly performing AI models for efficient resource allocation and explainable decision-making to ensure fairness, accountability, and compliance. To balance this trade off and inspired by the closed loop automation and XAI methodologies, this paper presents a novel explanation-guided \emph{in-hoc} federated learning (FL) approach where a constrained resource allocation model and an \emph{explainer} exchange---in a closed loop (CL) fashion---soft attributions of the features as well as inference predictions to achieve a transparent 6G network slicing resource management in a RAN-Edge setup under non-independent identically distributed (non-IID) datasets. In particular, we quantitatively validate the faithfulness of the explanations via the so-called attribution-based \emph{confidence metric} that is included as a constraint to guide the overall training process in the run-time FL optimization task. In this respect, Integrated-Gradient (IG) as well as Input $\times$ Gradient and SHAP are used to generate the attributions for our proposed in-hoc scheme, wherefore simulation results under different methods confirm its success in tackling the performance-explainability trade-off and its superiority over the unconstrained Integrated-Gradient \emph{post-hoc} FL baseline.
\end{abstract}

\vspace{5mm}

\section{Introduction}
\IEEEPARstart{6}{G} networks aim to support numerous simultaneous and diverse slices for various vertical use cases, resulting in increased complexity. Extensive research efforts have been dedicated to AI-based autonomous management and orchestration within the zero-touch network and service management (ZSM) framework, standardized by the ETSI, which aims to effectively handle the end-to-end network and services to accommodate the various services offered by 6G networks \cite{ZSM}. On the other hand, FL is a decentralized approach to machine learning that allows for the training of models on distributed data without the need to transfer the data to a central server.  The ZSM framework can incorporate FL as a key component for managing zero-touch distributed network slices while preserving the confidentiality of sensitive data \cite{ns3}. 
Furthermore, for the widespread adoption of AI in telecommunication networks, addressing the lack of transparency, trust, and explainability in black-box AI models becomes crucial \cite{trust}. The deployment of network automation requires a thorough understanding of AI models' decision-making and behavior. This drives the search for AI approaches that offer understandable and explainable outcomes.
In this regard, explainable AI (XAI) methods have emerged as a means to scrutinize the decisions made by black-box AI models. These methods build white-box models and generate feature contributions to explain AI decisions, promoting fairness, accuracy, and transparency. The ability to provide confidence and trust in AI models is vital for businesses and organizations deploying AI-enabled systems \cite{XAI,trust}. Nonetheless, there exists a clear trade-off between performance and explainability which is an ongoing challenge in the field of machine learning. Simpler models like linear regression or decision trees are more interpretable but may have limited predictive performance with complex data. On the other hand, complex models like deep neural networks can achieve higher accuracy but lack interpretability \cite{XAI}. Additionally, minimizing the model loss conflicts with its explainability.
In real-world deployments, it is essential to tackle the trade-off between performance and explainability towards ensuring trustworthy network orchestration. To achieve this goals, we anticipate that \emph{in-hoc} XAI approaches, that leverage the explanations during model's training and optimization, is a path to achieve that trade-off. We therefore introduce an explanation-guided \emph{in-hoc} FL strategy for achieving transparent zero-touch service management of 6G network slices at the RAN-Edge, specifically in a non-independent identically distributed (non-IID) data.
\subsection{Related Work}
This section discusses the state-of-the-art works on explainable AI (XAI) in the telecommunications domain. AI transparency is now more critical than ever with the advent of 6G networks, especially for zero-touch network management \cite{XAI,XAI2}. The "human-centric" character of 6G highlights how important it is for XAI to win over humans and keep them in the loop \cite{XAI2}. The authors in \cite{7} specify the challenges of developing XAI methods and show causal inference in 6G technology. 
The importance of explainability in 5G for essential services, device-to-device architectural security, and causal inference in 6G technology has also been emphasized  in \cite{21,XAI}. Moreover, The significance of explainability in upcoming 6G networks has been acknowledged by researchers \cite{ex1}, with applications to resolving handover and resource allocation issues. 
A study \cite{ex2} compares XAI techniques and suggests that SHAP is the best for identifying the cause of SLA violations. However, the study lacks clear explanations for the discrepancies among the results, casting doubt on the model's reliability.
Also, no quantitative metrics are used to evaluate the fidelity of the explanations. Thus, to resolve this gap, in \cite{rr}, an extensive range of XAI metrics is introduced for qualitative assessment, but no integration-based XAI methods are considered, and there is also no comparative analysis of different XAI techniques.
In \cite{XAI}, it is emphasized that the interpretability and transparency of AI/ML models are crucial for the full automation of a ZSM framework in 6G networks. Recently, \cite{neuro} proposes a neuro-symbolic XAI twin framework to enhance reasoning and trustworthiness in zero-touch IoE service management. However, they do not incorporate a  explainable-guided AI approach or address distributed learning with XAI in the ZSM framework like us. Additionally, in \cite{FL_xai2}, the concept of Federated Learning (FL) of XAI is introduced, focusing on its importance in 5G/6G networks. However, their work lacks quantitative validation of explanation faithfulness.
In \cite{XAI_FL}, the authors utilize the SHAP method from XAI to apply a trust-aware deep reinforcement learning-based device selection technique. This technique is employed in an FL training process for an autonomous driving scenario. On the other hand, several AI-based research works \cite{ns2,ns3} have focused on solving resource allocation issues in 6G network slices. However, most of these works have not addressed the challenge of providing explainability in their proposed solutions.
While some research areas like NLP and Healthcare have started recognizing the trade-off between model performance and explainability \cite{trade-off}, the field of telecommunication is still in its early stages concerning this concept. However, in a fortunate development, the authors \cite{XAI} have emphasized the importance of \emph{in-model} explainable AI (XAI) methods for the beyond 5G/6G field. They discuss the challenges and significance of achieving a better trade-off between explainability and AI model performance. They advocate for developing inherently explainable models and establishing quantifiable metrics to evaluate the effectiveness of XAI. \emph{In-model} explainability aims to create transparent models from the start, reducing reliance on post hoc XAI techniques. This approach allows stakeholders to understand the decision-making process or any decision while balancing performance and explainability.
In addition to the aforementioned research areas, a new research approach called Explanation Guided Learning (EGL) has emerged, which aims to exploit the explanations of machine learning models during the training to guide the model towards fulfilling both trustworthiness and performance. Specifically, EGL incorporates additional supervision signals or XAI-driven prior knowledge into the model's reasoning process, which helps improving the confidence of model's predictions \cite{EGL}, which might improve the balance between explainability and AI model performance.

In essence, the state-of-the-art work indicates that as 5G began to appear in 2020, discussions about the potential of 6G technology emerged. In addition to achieving higher bandwidth, enhanced reliability, energy efficiency, and lower latency, researchers see 6G as an advanced network that is "human-centric" and powered by AI. On the other hand, incorporating AI may result in intricate and opaque decision-making procedures. To tackle this issue, XAI has emerged as a promising solution in telecommunications.
However, there is a need for more comprehensive comparative analyses and quantifiable metrics for the evaluation of various XAI methods, along with a deeper exploration of emerging approaches like EGL in the context of telecommunications. Quantitative validation and insights into real-world deployment challenges are essential to bridge the gap between research and practical implementation. Addressing these gaps will be instrumental in achieving the desired balance between AI model performance and explainability, particularly in the evolving landscape of 6G networks.

\subsection{Contributions}
The main contributions of this paper are as follows:
\begin{itemize}
    \item We introduce a novel explanation-guided \emph{in-hoc} federated learning approach, where a constrained slice-level CPU resource allocation model and an \emph{explainer} exchange---in a closed loop way---explanations in the form of soft feature attributions as well as predictions to achieve a transparent and explainability-aware 6G network orchestration in a non-IID setup,
    
    \item We adopt the integrated gradients (IG) XAI method to generate explanations in terms of feature attributions, and map them to a soft probability space,
    
    \item These soft attributions are then used to quantitatively evaluating the model's \emph{confidence metric} which is included as a constraint in the FL optimization task.
    
    \item We formulate the corresponding XAI-constrained optimization problem under the \emph{proxy-Lagrangian} framework and solve it via a non-zero sum two-player game strategy, while comparing with a vanilla unconstrained post-hoc IG FL baseline.

    \item We present a comparative analysis of additional XAI methods that are used to generate attributions for the \emph{in-hoc} scheme and assess their confidence metrics via distribution plots. 
    
    \item We showcase the impact of network conditions (channel quality indicator (CQI), OTT traffics and MIMO full-rank usage) on the output CPU allocation.
\end{itemize}

\subsection{Notations}
We summarize the notations used throughout the paper in Table I.
\begin{table}[t!]
\label{Table1}
\centering	
\newcolumntype{M}[1]{>{\centering\arraybackslash}m{#1}}

\caption{\textcolor{black}{Notations}}
{\color{black}\begin{tabular}{m{2cm}M{6cm}}
\hline
\hline
\multicolumn{1}{>{\centering\arraybackslash}M{2cm}}{Notation} & \multicolumn{1}{>{\centering\arraybackslash}M{6cm}}{Description}\\
\hline
$S_\mu(\cdot)$ & Logistic function with steepness $\mu$\\
$F(\cdot)$ & CDF\\
$\tilde{F}(\cdot)$ & Complementary CDF\\
$L$ & Number of local epochs\\
$T$ & Number of FL rounds\\
$\mathcal{D}_{k,n}$ & Dataset at CL $(k,n)$\\
$D_{k,n}$ & Dataset size\\
$\ell(\cdot)$ & Loss function\\
$\mathbf{W}_{k,n}^{(t)}$ &  Local weights of CL $(k,n)$ at round $t$\\
$\mathbf{x}_{k,n}$ & Input features\\

$\hat{x}_{k,n}^{(i,j)}$ & Mutated features \\
$\hat{y}_{k,n}^{(i)}$ & Allocated resource by CL $(k,n)$\\
$\alpha_{n}$ & Resource lower-bound for slice $n$\\
$\beta_{n}$ & Resource upper-bound for slice $n$\\
$\nu_{n}$ & Threshold of Confidence bound for slice $n$ \\
$\mathcal{D}_{k,n}$ & Samples whose prediction fulfills the SLA \\

$\hat{z}_{k,n}^{(i)}$ & Predictions after mutation \\
$\pi_{k,n}^{(i,j)}$ & Probability Distribution\\
$C_{k,n}$ & Confidence Metric \\
$\alpha_{k,n}^{(i,j)}$ & Weighted attribution of features \\
$\lambda_{(\cdot)}$ & Lagrange multipliers\\
$R_\lambda$ & Lagrange multiplier radius\\
$\mathcal{L}_{(\cdot)}$ & Lagrangian with respect to ($\cdot$)\\
\hline
\hline
\end{tabular}}
\end{table}
\vspace{0.5cm}
\section{System Model and Problem Statement}
\subsection{System Model}

As depicted in Fig.~1, a 6G RAN-Edge topology under a per-slice central unit (CU)/distributed unit (DU) functional split is considered, wherein DUs are co-located with the transmission/reception point (TRP), while each CU $k$ is a virtual network function (VNF) running on top of a commodity hardware at the Edge domain and implements a closed loop (CL) $k\,(k=1,\ldots,K)$ consisting of a key performance indicators (KPIs) collection as well as AI-enabled slice resource allocation functions. This CL concept adheres to the ZSM framework ETSI \cite{ZSM}, which considers autonomous feedback loops between data monitoring, AI-driven data analytics, and decision-making functions to achieve particular network management tasks. In this regard, the architecture entails synthesizing data of several over-the-top (OTT) applications, which serves to build local datasets for each slice $n\,(n=1,\ldots,N)$, i.e., $\mathcal{D}_{k,n}=\{\mathbf{x}_{k,n}^{(i)},y_{k,n}^{(i)}\}_{i=1}^{D_{k,n}}$, where $\mathbf{x}_{k,n}^{(i)}$ stands for the input features vector, which includes OTT traffics, channel quality indicator (CQI) and multiple-input multiple-output (MIMO) full-rank usage, while $y_{k,n}^{(i)}$ represents the supervised output, which is the CPU load as shown in Table II. The considered slices include several OTTs each, namely,
\begin{itemize}
    \item \textbf{eMBB:} Netflix, Youtube and Facebook Video,
    \item \textbf{Social Media:} Facebook, Whatsapp and Instagram,
    \item \textbf{Browsing:} Apple, HTTP and QUIC,
\end{itemize}
where the corresponding accumulated datasets are non-IID due to the different traffic patterns induced by the heterogeneous users' distribution and channel conditions. Note that the intuitive link between the input features and the output resides in the fact that when the radio conditions are enhanced, the transmission and computing queues are relaxed, delivering thereby a higher number of packets which also results in increasing the CPU load.
Particularly, our system model incorporates MIMO technology, recognizing its pivotal role in enhancing data transmission and bolstering reliability through spatial diversity. Moreover, it introduces complexity into resource allocation decisions, particularly between CUs and DUs.
MIMO's adaptability to changing network conditions significantly impacts CPU load requirements, making it a critical factor in resource optimization. We combine MIMO-related insights with other relevant features to drive the optimization of resource allocation and enhance overall network performance.
\begin{figure}[t!]
\centering
    \includegraphics[scale=0.40, trim={4cm 0cm 0cm 4cm},clip]{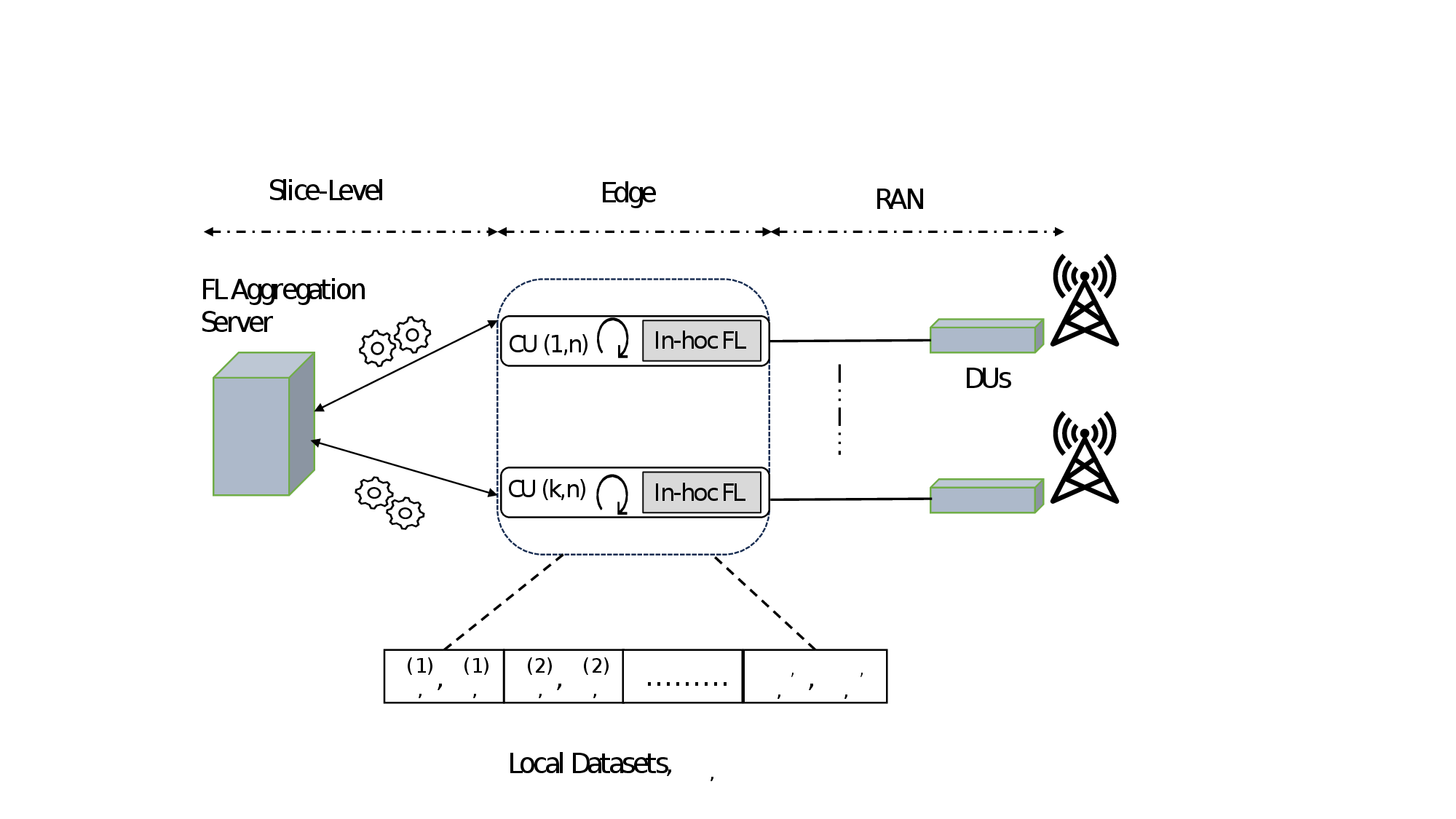}
\caption{Decentralized closed loops (CLs) architecture}
\end{figure}
Next, to enhance model accuracy and preserve data privacy as well as reduce transport overhead, the architecture adopts a federated learning (FL) approach, where local closed loops (CLs) participate in training using their respective synthesized datasets. The FL process is guided by XAI ensuring transparency and interpretability during model training. The knowledge gained from each local CL is then aggregated through an E2E slice-level server, collectively improving the global AI model's explainability and AI model performance.
\vspace{5mm}
\subsection{Problem Statement}
Unlike post-hoc XAI strategies, the aim of this paper is to design an explanation-guided \emph{in-hoc} FL-based resource allocation scheme for transparent zero-touch 6G network slicing at the RAN-Edge domain while balancing the trade-off between the performance and explainability. To achieve this, the proposed approach incorporates the XAI-based \emph{confidence metric} as constraint in a closed-loop manner during the model learning process. This allows for iterative optimization and adjustment of the model based on the insights gained from XAI considerations. Indeed, this approach is needed to strengthen the intelligence entities of our ZSM-based slice-level resource management \cite{neuro}.
Nevertheless, in our resource allocation scheme, we leverage feature attributions obtained through Integrated Gradients (IG) to understand the factors influencing CPU resource allocation for network slices. These attributions are mapped to a soft probability space, providing a probabilistic view of feature importance.
Notably, the model's confidence metric, derived from soft attributions, serves as a constraint in the allocation optimization. This approach yields transparent and explainability-aware resource allocation, aligning decisions with influential factors and the model's confidence. Moreover, we emphasize feature attributions, including those from MIMO technology, which plays a unique role in our system model, as discussed earlier. Insights from MIMO along with other features are integral to understanding factors guiding CPU resource allocation. This serve as the basis for our model's confidence metric, based on soft attributions, ensuring alignment with our goal of transparency and explainability in resource allocation.

\begin{table}[t!]
\label{Datasets-tab}
\centering	

\caption{Dataset Features and Output}
{\color{black}\begin{tabular}{lc}
\hline
\hline 

Feature & Description\\
\hline
\texttt{OTT Traffics per TRP}& \makecell{Apple, Facebook, Instagram,\\ NetFlix, HTTPS, Whatsapp,\\ and Youtube} \\ 
\texttt{CQI} & \makecell{Channel quality indicator \\ reflecting the average quality \\ of the radio link of the TRP} \\
\texttt{MIMO Full-Rank} & \makecell{Usage of MIMO full-rank \\spatial multiplexing in (\%)} \\

\hline 
\hline 
& \\
\hline
\hline 
Output & Description\\
\hline
\texttt{CPU Load} & CPU resource consumption (\%)\\
\hline
\hline
\label{Datasets-tab1}
\end{tabular}
}
\end{table}
\vspace{3mm}
\section{Proposed Approach}
Towards bridging the AI performance-explainability trade-off for trustworthy RAN slicing resource allocation, we detail in this section the proposed architecture shown in Fig. \ref{turbofl}.
\vspace{2mm}
\subsection{EGL-Driven Trustworthy Resource Allocation}
\begin{figure}[t!]
\centering
    \includegraphics[scale=0.80, trim={0 0 2mm 0},clip]{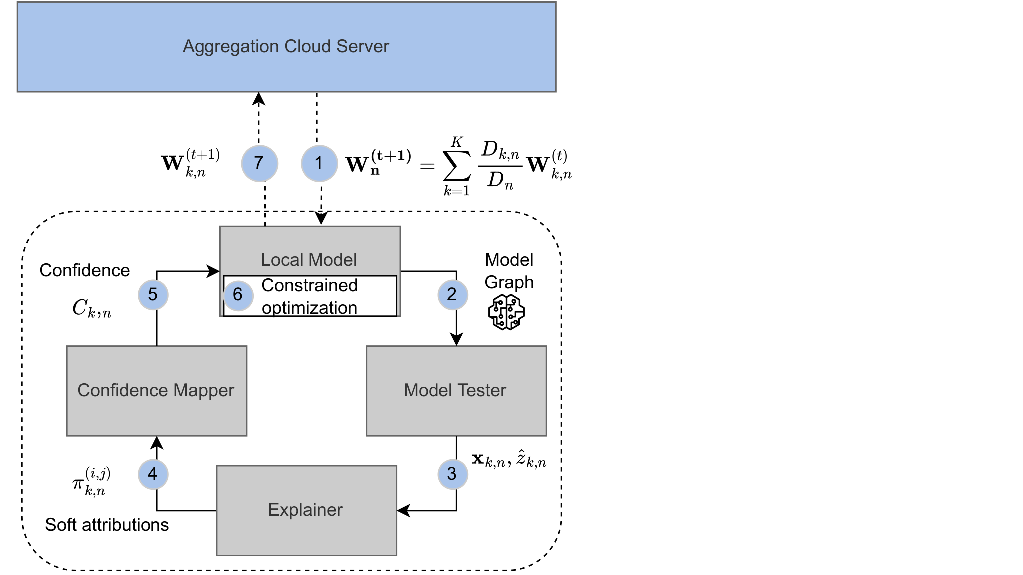}
\caption{\emph{in-hoc} FL for transparent resource allocation.}
\label{turbofl}
\end{figure}

Inspired by the CL and EGL principle, we propose an explanation-aided \emph{in-hoc} federated learning architecture where the local learning is performed iteratively with run-time explanation.
The overall working principle of the proposed model is manifested in Fig. \ref{turbofl} by clearly indicating the steps. For each local epoch, the Learner module feeds the posterior symbolic model graph to the Tester block which yields the test features and the corresponding predictions $z_{k,n}^{(i)}$ to the Explainer as shown from steps 1 to 3. Here, the model graph encapsulates the neural network's architecture (i.e., the arrangement of layers and their connectivity) and the learned parameters, which are the weights associated with each connection between neurons or nodes in the network. These weights are learned during training to make the model's predictions more accurate. The latter first generates the features attributions using one of the feature attribution XAI methods. It then converts these attributions to a soft probability distribution, as indicated in step 4, which is translated afterward into a confidence metric by the Confidence Mapper and fed back to the Learner to include it in the local constrained optimization, as pointed out in stages 5 and 6.
Indeed, for each local CL $(k,n)$, the predicted amount of resources
$\hat{y}_{k,n}^{(i)},\,(i=1,\ldots,D_{k,n})$, should minimize the main loss function with respect to the ground truth $y_{k,n}^{(i)}$ and guided by the gained insights from XAI. Hence, as depicted in those mentioned steps 1 to 7, the optimized local weights at round $t$, $\mathbf{W}_{k,n}^{(t)}$, are sent to the server which generates a global FL model for slice $n$ as,
\begin{equation}
\label{GlobalFL}
    \mathbf{W}_{n}^{(t+1)}=\sum_{k=1}^K \frac{D_{k,n}}{D_n}\mathbf{W}_{k,n}^{(t)},
\end{equation}
where $D_n=\sum_{k=1}^K D_{k,n}$ is the total data samples of all datasets related to slice $n$. The server then broadcasts the global model to all $K$ CLs that use it to start the next round of turbo local optimization. Specifically, it leverages a two-player game strategy to jointly optimize over the objective  and original constraints as well as their smoothed surrogates and detailed in the sequel.
\subsection{Model Testing and Explanation}
In this subsection, operational details of Model Tester and Explainer blocks of Fig. \ref{turbofl} are discussed more descriptively.
As depicted in stage 2 of Fig. 2, upon the reception of the updated model graph, the Tester uses a batch drawn from the local dataset to reconstruct the test predictions $\mathbf{z}_{k,n}^{(i)}$. All the graph, test dataset and the predictions are fed to the Explainer at stage 3 to generate the attributions, i.e., a quantified impact of each single feature on the predicted output. Let $\mathbf{a}_{k,n}^{(i)}\in \mathbb{R}^{Q}$ denote the attribution vector of sample $i$, which can be generated by any attribution-based XAI method. In our solution, for estimating the attributions of the considered features, we mainly leverage the low-complexity Integrated Gradient (IG) scheme \cite{IG}, which is based on the gradient variation when sampling the neighborhood of a feature. At stage 4, the Explainer then calculates what we call \emph{soft attributions} by mapping the attributions into a probability space as follows,
\begin{equation}
    \label{softAtt}
         {\pi_{k,n}^{(i,j)}= \frac{\exp \Bigl\{\abs{\alpha_{k,n}^{(i,j)}}\Bigr\}}{\sum _{q=1}^{Q}{\exp \Bigl\{\abs{\alpha_{k,n}^{(i,q)}}\Bigr\}}},\,j=1,\ldots,Q}, 
\end{equation}
where, $\alpha_{k,n}^{(i,j)}$ = $a_{k,n}^{(i,j)}/x_{k,n}^{(i,j)}$ stands for the weighted attribution, since the sensitivity of the model's output with respect to an input feature in the neighborhood of the input is approximately given by the ratio of the attribution for that input feature to the value of that feature \cite{con}.
\vspace{5mm}    
\subsection{Confidence Mapping}

To characterize the trustworthiness of the local model, we invoke the explanation-based confidence metric $C_{k,n}$ \cite{con}. Its rationale lies in the fact that slightly shifting the value of high-magnitude features (in the sens of attributions) is an acceptable way to determine the model's conformance. 
In essence, the explanation (feature attributions) informs the process of assessing confidence. When a model's explanation highlights certain features as highly influential, the confidence in the model's predictions can be gauged by examining how robust those predictions are when these significant features are altered in the input neighborhood. Therefore, the explanation (attributions) directly guides the method for evaluating and understanding the model's confidence in its predictions.
In our context, such an approach is vital because there will likely be no change in the SLA group (SLA violation or non-violation) if we sample over low attribution features space. 
In this respect, the \emph{Confidence Mapper} at stage 5 of Fig. 2 starts by performing a feature mutation, where it selects from the dataset feature $x_{k,n}^{(i,j)}$ with probability $\pi_{k,n}^{(i,j)}$, and changes it to the baseline value, i.e., zero, 
\begin{equation}
    \hat{x}_{k,n}^{(i,j)} = x_{k,n}^{(i,j)} \times (1 -  p), \,\,p \sim \mathcal{B}\left(1, \pi_{k,n}^{(i,j)}\right),
\end{equation}
to force the change of the model's prediction into the opposite category.
By adding Fig. 3, we have clarified the elucidation of the feature mutation process within the Confidence Mapper box of Fig. 2. This figure breaks the procedure into two steps, with numbered blue markers designating each step in order. In Stage (a), eligible features for mutation are identified, with the Feature Selector making these selections based on prior discussions.
In classification tasks, the categories (or subsets) are merely the classes. In contrast, we cast CPU resource allocation as a regression problem since it is well-suited for predicting continuous variables, allowing the FL model to estimate the CPU load as a numerical value. In this case, the subsets are defined according to an SLA threshold, i.e., 
\begin{equation}
    \mathcal{D}_{k,n}=\mathcal{U}_{k,n} \cup \bar{\mathcal{U}}_{k,n},
\end{equation}
where $\mathcal{U}_{k,n}$ contains the samples whose prediction fulfills the SLA, i.e., their CPU load lies in an interval $[\alpha_n,\beta_n]$.
The aforementioned transformation leads to a mutated dataset $\{\hat{x}_{k,n}^{(i,j)}\}$. The Confidence Mapper reports then the fraction of samples in the neighborhood for which the decision of the model,
\begin{equation}
    \hat{z}_{k,n}^{(i)}=\mathcal{M}_{k,n}(\mathbf{W}_{k,n}^{(t)}, \hat{\mathbf{x}}_{k,n}^{(i)}),
\end{equation}
does not move to the other set, that is, conforms to the original decision, as the conservatively estimated confidence measure \cite{con}, i.e.,
\begin{equation}
\label{GlobalFL}
C_{k,n} =\frac{1}{{u_{k,n}}}\sum_{i=1}^{u_{k,n}}  \max \Bigl\{\mathds{1}_{\mathbb{R}^{-}}\left(\alpha_n - \hat{z}_{k,n}^{(i)}\right) , \mathds{1}_{\mathbb{R}^{-}} \left( \hat{z}_{k,n}^{(i)} - \beta_n \right)\Bigl\},
\end{equation}
where $\hat{z}_{k,n}^{(i)}$ are the predictions after mutation, while $u_{k,n}$ stands for the size of the considered original category $\mathcal{U}_{k,n}$.
This process is visually depicted in stage (b) of Fig. 3. Here, the Confidence Analysis block, upon receiving the newly mutated dataset $\mathbf{D}_{1}$, computes the confidence metric and later shares it with the local model, contributing to its optimization process.

\begin{figure}[h!]
\centering
    \includegraphics[scale=0.75, trim={0cm 0cm 0cm 0cm},clip]{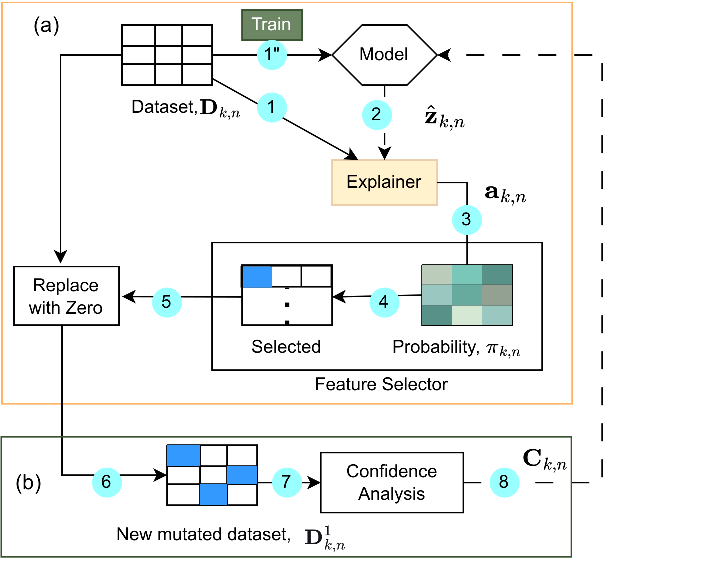}
\caption{Feature mutation approach of confidence mapper}
\end{figure}

\subsection{Explainability-Aware Resource Allocation}
We cast the slice-level resource allocation as a confidence-constrained regression task. For this purpose, we consider the datasets corresponding to the different slices as summarized in the Table II  and section II, where resources at CU-level are dynamically allocated to slices according to their traffic patterns and radio conditions (CQI and MIMO full-rank usage). In this respect, we formulate the constrained optimization using the proxy-Lagrangian framework and solve it via a non-zero sum two-player game strategy. 
Specifically, the confidence metric should be higher than a predefined threshold $\nu_n$. This translates into solving a statistically constrained local resource allocation problem as shown at stage 6 of Fig. \ref{turbofl}, i.e.,
\begin{subequations}
\label{OPT1}
\begin{equation}
    \min_{\mathbf{W}_{k,n}^{(t)}}\, \frac{1}{D_{k,n}}\sum_{i=1}^{D_{k,n}}\ell\left(y_{k,n}^{(i)}, \hat{y}_{k,n}^{(i)}\left(\mathbf{W}_{k,n}^{(t)},\mathbf{x}_{k,n}\right)\right),
\end{equation}

\begin{equation}
     C_{k,n} \geq \nu_{n}\label{Conf},
\end{equation}
\end{subequations}
which is solved by invoking the so-called \emph{proxy Lagrangian} framework \cite{Cotter}. This consists first on considering two Lagrangians as follows:
\begin{subequations}
\label{ProxyLagrangian}
\begin{equation}
\begin{split}
    \mathcal{L}_{\mathbf{W}_{k,n}^{(t)}}=&\frac{1}{D_{k,n}}\sum_{i=1}^{D_{k,n}}\ell\left(y_{k,n}^{(i)}, \hat{y}_{k,n}^{(i)}\left(\mathbf{W}_{k,n}^{(t)},\mathbf{x}_{k,n}\right)\right)
    \\& +\lambda_1\Psi_1\left(\mathbf{W}_{k,n}^{(t)}\right),
\end{split}
\end{equation}
\begin{equation}
    \mathcal{L}_{\lambda}=\lambda_1\Phi_1\left(\mathbf{W}_{k,n}^{(t)}\right)
\end{equation}
\end{subequations}
where $\Phi_{1}$ and $\Psi_{1}$ represent the original constraints and their smooth surrogates, respectively. Specifically, the indicator term in (\ref{Conf}) are replaced with Logistic functions and the soft maximum is used as surrogate of the maximum as,

\begin{equation}
\label{PRO11}
\begin{split}
    \Psi_1\left(\mathbf{W}_{k,n}^{(t)}\right)=\nu_{n} - \frac{1}{u_{k,n}}\sum_{i=1}^{u_{k,n}} & \log  \Bigl\{ \exp[S_{\mu}\left( \hat{z}_{k,n}^{(i)} - \alpha_n\right)]
    \\& + \exp[S_{\mu} \left( \beta_n - \hat{z}_{k,n}^{(i)}  \right)]\Bigl\}\leq 0,
\end{split}
\end{equation}
where $L_{\mu}$ stands for the Logistic function with steepness parameter $\mu$, i.e.,
\begin{equation}
    S_{\mu}\left(\theta\right)=\frac{1}{1+e^{-\mu\theta}}.
\end{equation}
\algblock{ParFor}{EndParFor}
\algnewcommand
\algorithmicparfor{\textbf{parallel for}}
\algnewcommand\algorithmicpardo{\textbf{do}}
\algnewcommand
\algorithmicendparfor{\textbf{end parallel for}}
\algrenewtext{ParFor}[1]{\algorithmicparfor\ #1\ \algorithmicpardo}
\algrenewtext{EndParFor}{\algorithmicendparfor}

\begin{algorithm}[t]
\caption{ \emph{In-hoc} Explainable Federated Learning}
\footnotesize
\SetAlgoLined

\KwIn{$K$, $m$, $\eta_{\lambda}$, $T$, $L$. \texttt{\# See Table III}}

Server initializes $\mathbf{W}_n^{(0)}$ and broadcasts it to the $\mathrm{CL}$s\\

\For{$t=0,\ldots,T-1$}{
\textbf{parallel for} $k=1,\ldots,K$ \textbf{do} \\
 Initialize $M=$ \texttt{num\_constraints} and $\mathbf{W}_{k,n,0}=\mathbf{W}_{n}^{(t)}$\\ 
 Initialize $\mathbf{A}^{(0)}\in \mathbb{R}^{(M+1) \times (M+1)}$ with $\mathbf{A}_{m',m}^{(0)}=1/(M+1)$\\
 \For {$l=0,\ldots,L-1$}{
 Receive the graph $\mathcal{M}_{k,n}$ from the local model\\
  \texttt{\# Test the local model and calculate the attributions}\\
  $a_{k,n}^{i,j}=$ \texttt{Int. Gradient} $\left(\mathcal{M}_{k,n}\left(\mathbf{W}_{k,n,l}, \mathbf{x}_{k,n}\right)\right)$\\
  \texttt{\# Generate soft attributions}\\
  $ {\pi_{k,n}^{(i,j)}= \frac{\exp \Bigl\{\abs{\alpha_{k,n}^{(i,j)}}\Bigr\}}{\sum _{p=1}^{L}{\exp \Bigl\{\abs{\alpha_{k,n}^{(i,p)}}\Bigr\}}},\,j=1,\ldots,Q}, $\\
  \texttt{\# Mutate the test dataset}\\
  Mutate $x_{k,n}^{(i,j)}$ with probability $\pi_{k,n}^{(i,j)}$\\
  \texttt{\# Calculate the confidence metric}\\
  $C_{k,n} = \frac{\sum_{i=1}^{u_{k,n}}  \max \Bigl\{\mathds{1}_{\mathbb{R}^{-}}\left(\alpha_n - \hat{z}_{k,n}^{(i)}\right) , \mathds{1}_{\mathbb{R}^{-}} \left( \hat{z}_{k,n}^{(i)} - \beta_n \right)\Bigl\}}{u_{k,n}}$\\
  Let $\lambda^{(l)}$ be the top eigenvector of $\mathbf{A}^{(l)}$\\
  \texttt{\# Solve problem (\ref{OPT1}) via oracle optimization}\\
  Let $\hat{\mathbf{W}}_{k,n,l}=\mathcal{O}_\delta\left(\mathcal{L}_{\mathbf{W}_{k,n,l}}(\cdot, \hat{\lambda}^{(l)})\right)$\\
  Let $\Delta_{\lambda}^{(l)}$ be a gradient of $\mathcal{L}_{\lambda}(\hat{\mathbf{W}}_{k,n,l}, \lambda^{(l)})$ w.r.t. $\lambda$\\
  \texttt{\# Exponentiated gradient ascent}\\
  Update $\tilde{\mathbf{A}}^{(l+1)}=\mathbf{A}^{(l)}\odot\cdot\exp{\eta_{\lambda}\Delta_{\lambda}^{(l)}(\lambda^{(l)})}$\\
  \texttt{\# Colunm-wise normalization}\\
  $\mathbf{A}_{m}^{(l+1)}=\tilde{\mathbf{A}}_{m}^{(l+1)}/\norm{\mathbf{A}_{m}^{(l+1)}}_1,\,m=1,\ldots,M+1$\\
  }
  \Return{$\hat{\mathbf{W}}_{k,n}^{(t)}=\frac{1}{L^{\star}}\sum_{l=0}^{L-1}\hat{\mathbf{W}}_{k,n,l}$}\\
  Each local CL $(k,n)$ sends $\hat{\mathbf{W}}_{k,n}^{(t)}$ to the server. \\
 \textbf{end parallel for}\\
 \Return{$\mathbf{W}_{n}^{(t+1)}=\sum_{k=1}^K \frac{D_{k,n}}{D_n}\hat{\mathbf{W}}_{k,n}^{(t)}$}\\
 and broadcasts the value to all local CLs.}
\end{algorithm}

This optimization task turns out to be a non-zero-sum two-player game in which the $\mathbf{W}_{k,n}^{(t)}$-player aims at minimizing $\mathcal{L}_{\mathbf{W}_{k,n}^{(t)}}$, while the $\lambda$-player wishes to maximize $\mathcal{L}_{\lambda}$ \cite[Lemma 8]{TwoPlayer}. While optimizing the first Lagrangian w.r.t. $\mathbf{W}_{k,n}$ requires differentiating the constraint functions $\Psi_1(\mathbf{W}_{k,n}^{(t)})$, to differentiate the second Lagrangian w.r.t. $\lambda$ we only need to evaluate $\Phi_1\left(\mathbf{W}_{k,n}^{(t)}\right)$. Hence, a surrogate is only necessary for the $\mathbf{W}_{k,n}$-player; the $\lambda$-player can continue using the original constraint functions. The local optimization task can be written as,
\begin{subequations}
\label{ProxyLagrangian}
\begin{equation}
\begin{split}
    \min_{\mathbf{W}_{k,n}\in \Delta} \,\,\,\,\max_{\lambda,\, \norm{\lambda}\leq R_\lambda}\,\,\mathcal{L}_{\mathbf{W}_{k,n}^{(t)}}
\end{split}
\end{equation}
\begin{equation}
        \max_{\lambda,\, \norm{\lambda}\leq R_\lambda}\,\,\,\,\min_{\mathbf{W}_{k,n}\in \Delta} \mathcal{L}_{\lambda},
\end{equation}
\end{subequations}
where thanks to Lagrange multipliers, the $\lambda$-player chooses how much to weigh the proxy constraint functions, but does so in such a way as to satisfy the original constraints, and ends up reaching a nearly-optimal nearly-feasible solution \cite{Gordon}. These steps are all summarized in Algorithm 1.

At line 1 of Algorithm 1, we declare the baseline parameters. Then, at line 2, the aggregation server initializes the weights and broadcasts them to all participating CLs. Lines 3 to 4 describe the execution of the FL optimization task, as mentioned earlier at each FL round.Noted that, the preference of proxy-Lagrangian optimization over the well know convex-concave procedure \cite{ccp} is firmly justified. This decision is primarily driven by the intricacy of the non-convex, dataset-dependent XAI constraints, which CCP may not accurately capture. Moreover, it provides computational efficiency, ensuring timely resource allocation decisions, unlike CCP, which often involves computationally expensive iterations for convexification. The versatility and numerical stability of proxy-Lagrangian methods make them well-suited for handling a wide range of SLA constraints and statistical metrics. In summary, proxy-Lagrangian optimization outperforms CCP in terms of efficiency and flexibility  in addressing B5G resource allocation challenges.
\vspace{2mm}
\section{Results}
In this section we evaluate the proposed \emph{in-hoc} FL framework by first justifying the use of feature attributions as a pillar to build the explainability-aware constrained resource allocation model. We then present the FL convergence, confidence score and performance metrics and showcase their underlying trade-offs. We afterward study the correlation between features attributions and draw some important conclusions. Finally we present a time complexity analysis  to assess the computational efficiency of the proposed framework.To implement the model Tester and Explainer, we invoke \texttt{DeepExplain} framework, which includes state-of-the-art gradient and perturbation-based attribution methods \cite{Deep}. It provides an attribution score based on the feature's contribution to the model's output, which we integrate with our proposed constrained resource allocation FL framework in a closed-loop iterative way. Besides, our simulation-based experiments were conducted in a Python environment on Ubuntu 20.04.  The proposed federated neural network was built with TensorFlow V1, featuring fully connected layers activated by the leaky ReLU function. We used the Adam optimizer with a learning rate $\eta_{\lambda}$ of 0.02 to update the model's parameters.
\vspace{2mm}
\subsection{Parameter Settings and Baseline}
Here, we define the parameters to tackle the FL optimization task. The details of the considered slices and associated KPIs are mentioned in section II.
We use vectors $\alpha$, $\beta$ for the cpu resource bounds corresponding to the different slices for a particular resource and $\nu$ for the explainability confidence metrics threshold. The parameters settings are presented in Table III.
Recently, there has been growing interest in combining FL and XAI approaches. In this work, the considered baseline is the state-of-the-art unconstrained FL with post-hoc explanation \cite{baseline_new}. The authors \cite{baseline_new} investigate the feature importance challenge in vertical FL scenarios and propose a method, the Sharp Federated algorithm, which utilizes Shapley values to determine feature importance in a post-hoc way.
\begin{table}[t!]
\label{Table1}
\centering	
\newcolumntype{M}[1]{>{\centering\arraybackslash}m{#1}}
\caption{Settings}
\begin{tabular}{ccc}
\hline 
\hline

Parameter & Description & Value\\
\hline
$N$ & \# Slices & $3$\\
$K$ & \# CLs & $100$ \\ 
$D_{k,n}$ & Local dataset size & $1000$ samples\\ 
$\alpha_n$ & SLA lower bound & $0$\%\\ 
$\beta_n$ & SLA upper bound & $3$\%\\ 
$T$ & \# Max FL rounds & $30$\\  
$L$ & \# Local epochs & $100$\\ 
$R_{\lambda}$ & Lagrange multiplier radius &  Constrained: $10^{-5}$\\
$\eta_{\lambda}$ & Learning rate & $0.02$\\ 
\hline
\hline
\label{FLsettings}
\end{tabular}
\end{table}

\subsection{Bridging the Performance-Explainability Trade-off}
\begin{figure}[t]
    \centering
    \includegraphics[width=0.42\textwidth,height=4.2cm]{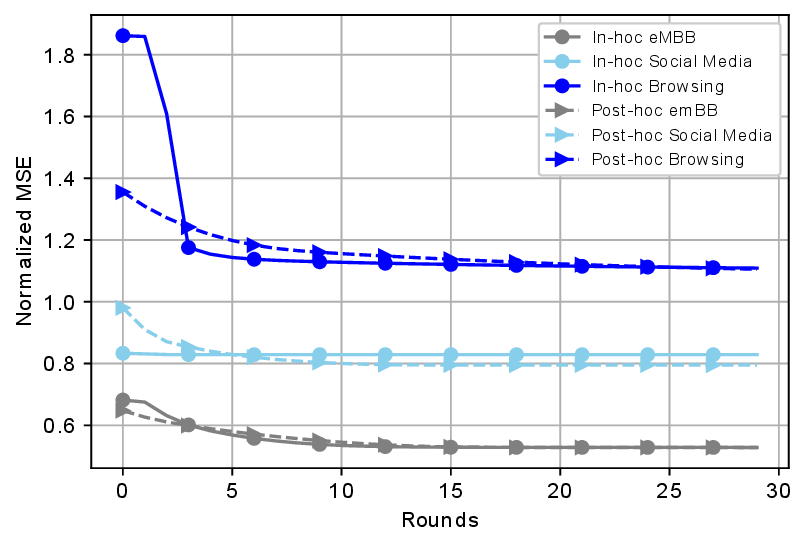}
    \caption{FL training MSE loss vs. FL rounds with $\nu=[0.82, 0.83, 0.85]$}
    \label{convergence_plot}
    \vspace{-2mm}
\end{figure}

\begin{figure}[t]
    \subfloat[\centering Model confidence metric vs. FL rounds with slices' confidence bounds $\nu= {[}0.82, 0.83, 0.85{]}$.]{\includegraphics[width=0.42\textwidth,height=4.2cm]{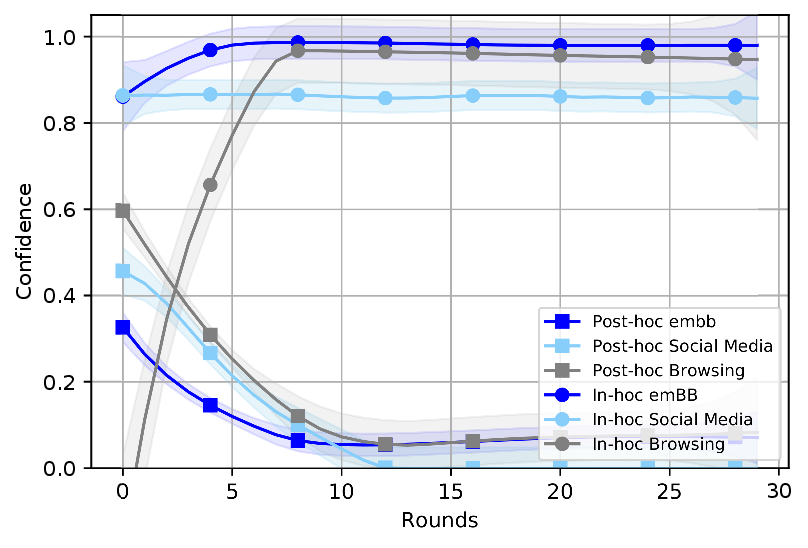}}\hspace{1.2cm}
    \qquad 
    \subfloat[\centering Confidence metric with the proposed in-hoc scheme on a test dataset.]{
        \includegraphics[width=0.42\textwidth,height=4.2cm]{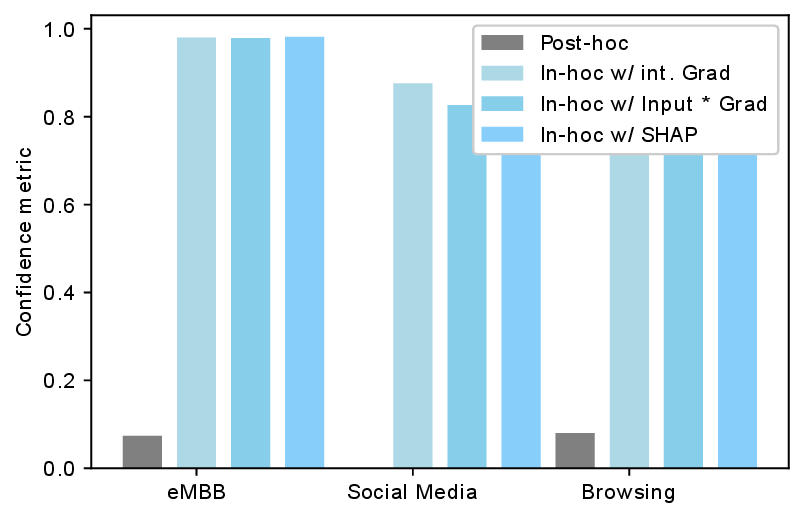}}
        \caption{Confidence metric using different XAI methods.}
        \label{confidence_plot}
\end{figure}

To study the performance-explainability trade-off of the proposed strategy, we plot both the convergence curves and the confidence metric vs. the FL rounds in Fig. \ref{convergence_plot} and Fig. \ref{confidence_plot}, respectively, considering our \emph{in-hoc} FL scheme and the baseline unconstrained IG post-hoc. It shows that the confidence metric of  the proposed in-hoc approach for the different slices conserves a higher value above 85$\%$, while presenting a similar convergence trend as the post-hoc FL baseline. In contrast, the latter fails to ensure the confidence of the model, since its confidence metric decreases as we gradually approach the convergence. This behavior conveys that the in-hoc strategy addresses the trade-off successfully, and guarantees explainability and trust in the training phase. For analysis completeness, the Explainer block in Fig. \ref{turbofl} is also implemented using both the perturbation-based XAI method SHAP and gradient-based method Input$\times$Gradient \cite{com_XAI}, where Fig. \ref{confidence_plot}-(b) confirms that our proposed \emph{in-hoc} FL algorithm preserves the same behavior during the testing phase. Moreover, the confidence score has remained almost the same even if the attribution scores have been generated by various XAI methods, which makes our proposed algorithm more reliable.

Overall, based on the presented results in Fig. \ref{convergence_plot} and Fig. \ref{confidence_plot}, the constrained \emph{in-hoc} FL ensures a trade-off between convergence and confidence, i.e.,  while the model loss decreases within an allowable confidence threshold, the model's confidence increases and significantly outperforms the post-hoc baseline. This indicates that the model becomes more confident in its predictions as it converges. This is achieved thanks to the explanation-guided \emph{in-hoc} constrained FL optimization. On the contrary, in the state-of-the-art post-hoc scenario, the model confidence degrades as it starts to converge.

 \vspace{-0.5mm}

\subsection{Explaining the Impact of Network Parameters on Slice Resource Allocation}
\begin{figure}[h]
        \centering
        \includegraphics[width=0.47\textwidth,height=3.8cm]{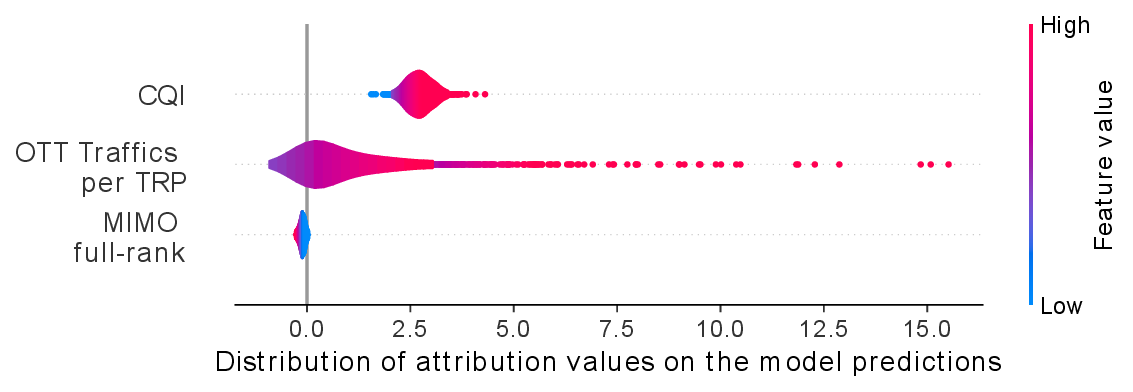}
        \caption{Distribution plot of attribution for eMBB slice}
        \label{shap_values}
        \vspace{-2mm}
\end{figure}

In this subsection, we first examine the post-hoc feature attributions plots of the eMBB slice, which are generated via SHAP as illustrated in Fig \ref{shap_values}. The analysis reveals that CQI has the highest impact on the CPU allocation, as it is widely concentrated towards positive values. This concentration indicates that better channel quality is leading to lower CPU load, which can be interpreted by the reduction of retransmissions and queuing time. Consequently, the model can anticipate efficient data transmission, increased network capacity, and potential resource optimization. Additionally, the concentration of OTT traffic per TRP around the lowest positive values highlights varying traffic levels that impact CPU load. Lower levels of OTT traffic typically require less processing resources, leading to decreased CPU load. The FL model can also leverage this information to optimize resource allocation and guide network planning and scaling efforts. Furthermore, the negative concentration of MIMO full rank suggests signal degradation and interference, which could potentially lead to higher CPU load predictions. However, in our specific case, where the value is very small, the impact of MIMO full rank on CPU load can be considered negligible. Nonetheless, this analysis still allows the FL model to identify and address potential issues related to signal degradation and interference, providing valuable insights for network optimization in other scenarios where MIMO full rank plays a more significant role. By considering these features, the FL model can make accurate predictions and enable resource optimization, leading to improved network performance and enhanced management of CPU load in network slices in a transparent and explainable way.

\subsection{Time complexity \& overhead}
\begin{figure}[t!]
        \centering
        \includegraphics[width=0.42\textwidth,height=4.5cm]{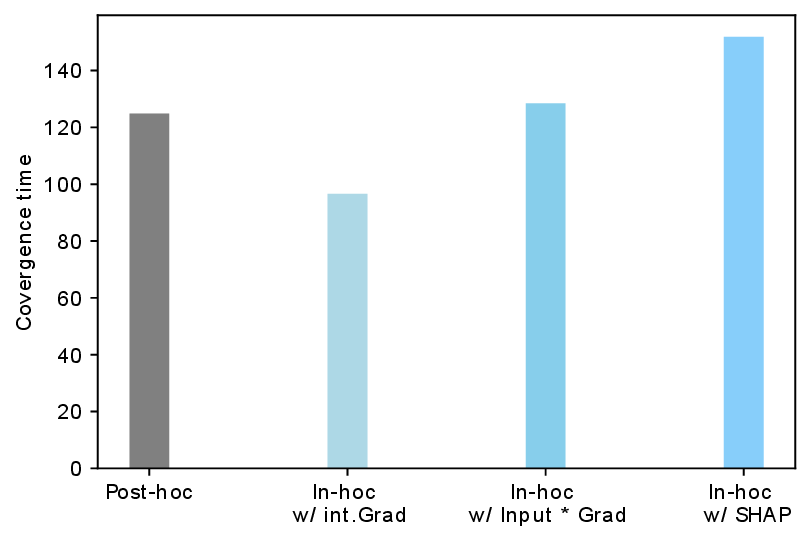}
        \caption{Convergence time of the proposed \emph{in-hoc} FL under different explainers vs. the post-hoc baseline.}
     \label{complexity}   
\end{figure}

Attention has been also paid to the training time having in mind future actual deployment of our solution. The efficiency of the solution's training process directly impacts its real-world usability and effectiveness. Faster training times enable the system to respond promptly to dynamic changes, ensuring real-world efficiency and cost-effectiveness. Additionally, efficient training times facilitate scalability, allowing the system to handle increasing data volumes and user demands without compromising performance. For evaluating the time complexity of the proposed solution, the \emph{in-hoc} FL explainer can be implemented with various feature attribution generators, we use IG as the primary method. Fig. \ref{complexity} shows the convergence time of the browsing slice under our \emph{in-hoc} FL strategy with SHAP, IG and Input $\times$ Grad as attribution scores generators as well as the baseline unconstrained post-hoc method. Since Fig. \ref{convergence_plot} indicates that the browsing slice with \emph{in-hoc} FL starts to converge at around round 13, the computation time has been calculated until that convergence round. It can observe from the Fig. \ref{complexity} that \emph{in-hoc} FL with IG takes lower time than others but nearly close to \emph{post-hoc} one while presenting the highest confidence.
This finding suggests that utilizing IG as the primary feature attribution method in the \emph{in-hoc} FL explainer enables faster convergence without compromising the confidence of the predictions. This is a significant advantage for real-world deployment, as it indicates efficient allocation of computational resources while maintaining the trustworthiness of the model.

\begin{table}[h!]
\label{overhead}
\centering	

\caption{Overhead comparison (KB) for Browsing slice}
{\color{black}\begin{tabular}{ccc}
\hline
\hline 

Scheme & Convergence round & Overhead(KB) \\
\hline

\texttt{In-hoc} &   13 & 16510\\ 
\texttt{Post-hoc}& 20 &  77714 \\ 
\hline 
\hline 
& \\
\label{overhead}
\end{tabular}
}
\end{table}
\vspace{-5mm}

Moreover, the results of Table IV reveal that the complexity introduced by XAI has a notable impact on the overall system performance. A significant overhead disparity can be observed in comparing the proposed \emph{in-hoc} FL and the baseline \emph{Post-hoc} FL. The \emph{in-hoc} FL method exhibited substantially lower overhead (16,778 KB) at the outset of convergence in round 12, in contrast to the \emph{Post-hoc} FL (77,776 KB), which commenced convergence at round 17. This substantial difference suggests that the \emph{in-hoc} FL approach offers notable advantages regarding communication efficiency. Notably, overhead is directly proportional to energy consumption. Hence, it is reasonable to infer that the \emph{in-hoc} FL method likely entails a lower computational load and potentially higher energy efficiency due to minimized data exchange.
Specifically, the reduction in overhead for the \emph{in-hoc} FL method suggests that the complexity introduced by XAI in this context leads to improved communication efficiency, which translates into potential benefits in computational load and energy efficiency. Such an approach would provide a more holistic understanding of the performance trade-offs associated with XAI complexity.
In summary, our analysis provides critical insights into our solution's complexity and overhead. The in-hoc FL approach with IG achieves a 16.67$\%$ reduction in computation time compared to the Post-hoc FL, ensuring faster convergence without compromising prediction confidence, which is vital for real-world adaptability.  Additionally, we observed a significant reduction in overhead with our in-hoc FL approach, signifying improved communication efficiency and potential gains in computational load and energy efficiency. This balanced approach positions our solution as a strong prospect for real-world deployment, offering efficiency and trustworthiness.

 \vspace{-1mm}
\subsection{Scalability analysis}
\vspace{-1mm}

\begin{figure}[t!]
        \centering
        \includegraphics[width=0.45\textwidth]{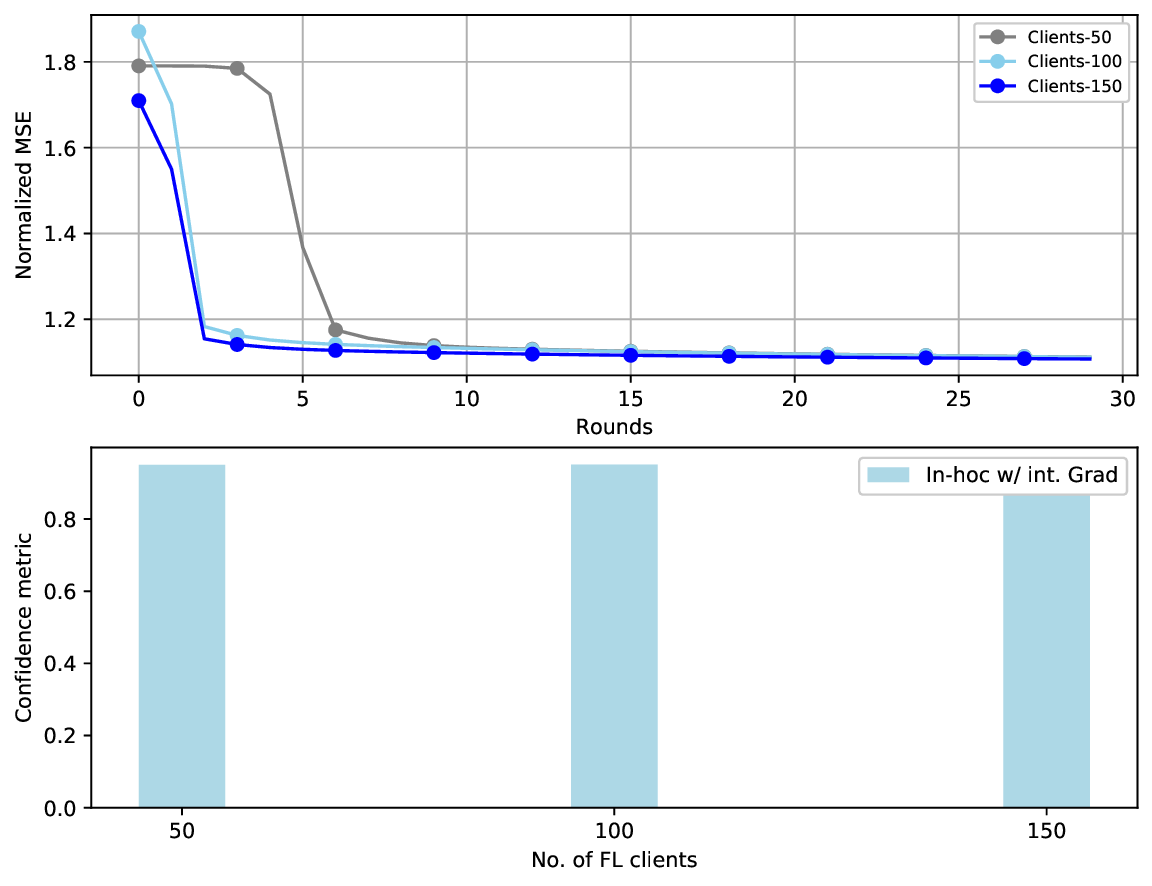}
        \caption{Scalability of the proposed In-hoc FL approach}
     \label{scalability} 
     \vspace{-5mm}
\end{figure}

Scalability is paramount in FL, ensuring consistent system performance regardless of the number of participating clients.
Fig. \ref{scalability} shows that our \emph{in-hoc} FL approach exhibits robust scalability. Firstly, it consistently exhibits stable convergence behavior, irrespective of changes in the number of participating clients. Secondly, the confidence metric, essential for assessing the prediction reliability, consistently remains above the 90$\%$ threshold within our in-hoc federated learning approach. This adaptability ensures reliable performance in dynamic real-world situations, making our FL solution well-suited for diverse deployment scenarios.

\section{Conclusion}
In this paper, we have presented a novel explanation-guided \emph{in-hoc} federated learning approach to achieve a transparent zero-touch service management of 6G network slices while bridging the trade-off between performance and explainability. We have considered the XAI-based confidence metric in a closed loop way to solve the underlying joint optimization task using a proxy-Lagrangian two-player game strategy. In particular, we have used both integrated-gradient and perturbation-based XAI schemes to generate the attributions consumed by our \emph{in-hoc} FL, which present almost the same superior performance compared to an unconstrained \emph{post-hoc} FL baseline. We have also provided a post-hoc analysis of the impact of networks parameters on the slice-level resource allocation, which points out that CQI is the key influencing feature. Finally, computational complexity (up to the convergence round) has been assessed which demonstrates that our \emph{in-hoc} FL has less complexity than the \emph{post-hoc} while presenting a clearly superior confidence. 
In the future, we aim to enhance model explainability through symbolic reasoning, which employs explicit rules and logic to make the decision-making process more understandable and trustworthy. Besides, we plan to contribute to benchmarking and standards, conduct real-world testing in 6G networks, explore user-centric and energy-efficient solutions, and ensure interoperability with existing technologies. These initiatives will improve the effectiveness and practicality of our approach in the dynamic 6G network landscape.
\section{Acknowledgment}
This work has been supported in part by the projects 6G-BRICKS (101096954) HORIZON-JU-SNS-2022 and ADROIT6G (101095363) HORIZON-JU-SNS-2022. Additionaly, this work was partially funded by MINECO (Spain) and the EU NextGenerationEU/PRTR (Call UNICO I+D 5G 2021, ref. number TSI-063000-2021-10)

\bibliographystyle{IEEEtran}
\bibliography{myBibliographyFile}

\balance

\end{document}